\documentclass[runningheads]{llncs}

\usepackage{eccv}

\usepackage{eccvabbrv}

\usepackage{graphicx}
\usepackage{booktabs}

\usepackage[accsupp]{axessibility}  

\usepackage{array}

\usepackage{hyperref}

\usepackage{orcidlink}

\begin{document}

\title{AIM 2024 Challenge on Video Super-Resolution Quality Assessment: Methods and Results} 

\titlerunning{AIM 2024 Challenge on Video Super-Resolution Quality Assessment}

\author{
Ivan~Molodetskikh\inst{1}\orcidlink{0000-0002-8294-0770}
\and Artem~Borisov\inst{1}\orcidlink{0009-0000-6124-885X}
\and Dmitriy~Vatolin\inst{1,2}\orcidlink{0000-0002-8893-9340}
\and Radu~Timofte\inst{3}\orcidlink{0000-0002-1478-0402}
\and Jianzhao~Liu\inst{4}
\and Tianwu~Zhi\inst{4}
\and Yabin~Zhang\inst{4}
\and Yang~Li\inst{4}
\and Jingwen~Xu\inst{4}
\and Yiting~Liao\inst{4}
\and Qing~Luo\inst{5}
\and Ao-Xiang~Zhang\inst{5,6}
\and Peng~Zhang\inst{5}
\and Haibo~Lei\inst{5}
\and Linyan~Jiang\inst{5}
\and Yaqing~Li\inst{5}
\and Yuqin~Cao\inst{7}
\and Wei~Sun\inst{7}
\and Weixia~Zhang\inst{7}
\and Yinan~Sun\inst{7}
\and Ziheng~Jia\inst{7}
\and Yuxin~Zhu\inst{7}
\and Xiongkuo~Min\inst{7}
\and Guangtao~Zhai\inst{7}
\and Weihua~Luo
\and Yupeng~Z.\inst{8}
\and Hong~Y.\inst{8}
\thanks{I.~Molodetskikh (\email{ivan.molodetskikh@graphics.cs.msu.ru}), A.~Borisov (\email{artem.\allowbreak borisov@\allowbreak graphics.cs.msu.ru}), D.~Vatolin (\email{dmitriy@graphics.cs.msu.ru}), and R.~Timofte (\email{radu.timofte@uni-wuerzburg.de}) were the challenge organizers, while the other authors participated in the challenge. \Cref{affiliations} contains the authors’ teams and affiliations. AIM 2024 webpage: \url{https://www.cvlai.net/aim/2024/}}
}

\authorrunning{I.~Molodetskikh, A.~Borisov, D.~Vatolin, R.~Timofte et al.}

\institute{Lomonosov Moscow State University, Russia
\email{\{ivan.molodetskikh,artem.borisov,dmitriy\}@graphics.cs.msu.ru}
\and MSU Institute for Artificial Intelligence, Russia
\and University of Würzburg, Germany
\and ByteDance, China
\and Tencent, China
\and School of Computer Science and Cyber Engineering, Guangzhou University, China
\and Shanghai Jiao Tong University, Shanghai, China
\and Ricoh Software Research Center Beijing, China
}

\maketitle

\begin{abstract}
  This paper presents the Video Super-Resolution (SR) Quality Assessment (QA) Challenge that was part of the Advances in Image Manipulation (AIM) workshop, held in conjunction with ECCV 2024.
  The task of this challenge was to develop an objective QA method for videos upscaled 2\texttimes{} and 4\texttimes{} by modern image- and video-SR algorithms.
  QA methods were evaluated by comparing their output with aggregate subjective scores collected from >150,000 pairwise votes obtained through crowd-sourced comparisons across 52 SR methods and 1124 upscaled videos.
  The goal was to advance the state-of-the-art in SR QA, which had proven to be a challenging problem with limited applicability of traditional QA methods.
  The challenge had 29 registered participants, and 5 teams had submitted their final results, all outperforming the current state-of-the-art.
  All data, including the private test subset, has been made publicly available on the challenge homepage at \url{https://challenges.videoprocessing.ai/challenges/super-resolution-metrics-challenge.html}.
  \keywords{Video Super-Resolution \and Quality Assessment \and Challenge}
\end{abstract}

\section{Introduction}
\label{sec:intro}

As consumer devices continue to increase in screen resolution, the task of image- and video-upscaling remains among the top research topics in the field.
In 2024 alone, several novel video Super-Resolution (SR) methods had appeared~\cite{kai2024evtexture,zhou2024video,li2024cfdvsr}.
This rapid development pace elevates the need for accurate objective Quality Assessment (QA) methods for super-resolved images and videos.

Current SR research commonly uses established image QA metrics, such as PSNR, SSIM~\cite{1284395}, and LPIPS~\cite{8578166}.
However, recent benchmarks~\cite{videometricsbenchmark,srmetricsbenchmark} show that these metrics correlate poorly with human perception, especially when applied to SR-upscaled output.
In particular, the classical PSNR and SSIM methods, despite wide use in SR research papers, are unfit for accurately estimating SR quality.
Other, deep-learning-based, methods can have challenges capturing specific artifacts arising from SR methods.

Hence, the task of super-resolution quality assessment is different from the general task of image and video quality assessment.
Accurate evaluation of SR results requires metrics specifically tuned for this task.
Several such methods had appeared in recent years~\cite{kirillova2022erqa, ma2017nrqa} and show promising results.
However, an accuracy gap remains between subjective evaluation and even the state-of-the-art SR QA metrics.

To help advance the SR QA metrics research, we organized a challenge for video super-resolution quality assessment, jointly with the Advances in Image Manipulation (AIM) 2024 workshop.
Participants develop an objective QA metric, that is then evaluated on 1200 videos upscaled by 52 modern SR methods.
The ground-truth scores are aggregated from >150,000 pairwise crowd-sourced votes.
The videos are divided into three difficulty levels, based on the behavior of existing QA metrics.

In the following sections we describe the challenge in more detail, and show the results and the participants' proposed approaches.

\section{Related Work}
\label{sec:related}

Video Super-Resolution (VSR) aims at restoring High-Resolution (HR) videos from their Low-Resolution (LR) counterparts.
It has extensive applications in various domains such as surveillance, virtual reality, and video enhancement.
This topic is actively evolving, with state-of-the-art approaches changing every year~\cite{pwc_vsr}.

VSR methods are prone to producing specific artifacts, which makes it challenging to evaluate their quality using the standard metrics, such as PSNR or SSIM.
Therefore, several SR-specific metrics appeared in recent years.


\subsection{Super-Resolution Quality Assessment}

Ma et al.~\cite{ma2017nrqa} proposed to use statistics computed from spatial and frequency domains to represent SR images.
Each set of extracted features is trained in separate ensemble regression trees, and a linear regression model is used to predict a quality score of image.
Unfortunately, on real data this metric performs worse than many current SOTA approaches in image- and video-QA.
In addition, it is non-differentiable, which limits its applicability for fine-tuning VSR models.

The main idea of ERQA~\cite{kirillova2022erqa} is to correctly assess an SR method's performance on edge-restoration.
This metric uses the Canny algorithm to find edges on distorted and ground-truth frames.
Then, $F_1$-score is used to compare them and compute the final score of the frame.
This method performs well for low-resolution video, but shows worse results for high-resolution video.
It is also non-differentiable.


\subsection{General Quality Assessment}

We also overview some approaches to the general image- and video-QA task that proved to work well for the SR QA case.

The main ideas of PieAPP~\cite{Prashnani_2018_CVPR} are the use of pairwise learning (which makes the process of image evaluation by the metric most similar to the process of subjective evaluation using the Bradley-Terry model~\cite{bradley1952rank}), and training on images with a big number of distortions (75 different distortions, including SR).
Learning on such a wide variety of artifacts likely contributed to this metric's excellent performance for SR QA task, according to benchmarks~\cite{srmetricsbenchmark}.

Q-Align~\cite{wu2023qalign} has become the SOTA method on many image- and video-QA and aesthetic assessment datasets.
It also shows excellent results among no-reference metrics in SR QA task.
This metric uses a multi-modal large language model mPLUG-Owl2~\cite{ye2023mplugowl2}, based on LLaMa-2-7B, to encode information about the image or video, as well as language instructions.
The model outputs probabilities for several quality levels: bad, poor, fair, good, and excellent.
These probabilities are combined to get the final score.
Training for three tasks at once allowed this method to become state-of-the-art on 12 datasets.

\section{AIM 2024 Video Super-Resolution QA Challenge}
\label{sec:challenge}

This challenge is one of the AIM 2024 Workshop\footnote{\url{https://www.cvlai.net/aim/2024/}} associated challenges on: sparse neural rendering~\cite{aim2024snr, aim2024snr_dataset}, UHD blind photo quality assessment~\cite{aim2024uhdbpqa}, compressed depth map super-resolution and restoration~\cite{aim2024cdmsrr}, raw burst alignment~\cite{aim2024rawburst}, efficient video super-resolution for AV1 compressed content~\cite{aim2024evsr}, video super-resolution quality assessment, compressed video quality assessment~\cite{aim2024cvqa} and video saliency prediction~\cite{aim2024vsp}.

We started the development phase on May~31st, and the final test phase on July~24th.
A total of 29 participants had registered for the challenge, 16~participants had sent in intermediate results, and 5 teams had submitted their code and results for the final evaluation.

\subsection{Challenge Goal}

The main objective of this challenge is to stimulate research and advance the field of QA metrics oriented specifically to super-resolution.
The task is to develop a state-of-the-art image/video quality assessment metric that correlates highly with subjective scores for super-resolved videos.

Concretely, participants were provided with video clips upscaled with a set of SR models.
The clips had different resolution ranging from 200\texttimes170 to 960\texttimes544.
The participants' metric had to produce a quality score for every video.
These quality scores were then compared against subjective scores obtained through crowd-sourced pairwise comparisons.
The following sections go into more detail about our data collection and submission evaluation processes.

Our challenge offered the participants both a Full-Reference (ground-truth high-resolution frames available) and a No-Reference (NR; only super-resolved frames available) track.
However, we only received submissions for the NR track.
We welcome this, as No-Reference quality assessment, while more challenging, has much wider applications.

\subsection{Dataset}
\label{sec:dataset}

We provide the participants with train (568 videos) and test (183 videos in public set and 373 videos in private set) subsets that cover many video super-resolution use-cases.
Train and public test subsets additionally had ground-truth subjective ranks available to participants, with the private test subset ground-truth ranks withheld until the end of the competition.
This section describes our process for collecting these videos and ground-truth ranks.

Most source videos come from the MSU Codecs Comparison~\cite{msu_cc} dataset.
These are high-bitrate open-source videos from the \url{https://vimeo.com} video hosting.
We split them into 10 clusters, and select one video closest to the center of each cluster, based on following metrics' values: Noise Estimation and Blurring from MSU VQMT~\cite{vqmt}, DBCNN~\cite{8576582}, and ClipIQA+~\cite{wang2022exploring}.
We started with 30 metrics, but by removing each and observing changes in the clusters, we determined that this set of four metrics is complete.

The next step was to downsample each of the 10 videos by 2\texttimes{} and 4\texttimes{} using bicubic interpolation, and compress it with libx264, libx265 (500~kb/s, 1000~kb/s, and 2000~kb/s bitrates), and SVT-AV1 (quality parameter 40 and 60).
We then upscaled these compressed videos using the following super-resolution methods, chosen based on the types and strength of artifacts they generate.
\begin{itemize}
    \item Real-ESRGAN~\cite{wang2021realesrgan}: 2\texttimes{} and 4\texttimes{} models
    \item RealSR~\cite{Ji_2020_CVPR_Workshops}: 4\texttimes{}; DF2K and DF2K JPEG presets
    \item RVRT~\cite{liang2022rvrt}: 4\texttimes{}; REDS, Vimeo + BD and Vimeo + BI presets
    \item SwinIR~\cite{liang2021swinir}: 2\texttimes{} default model, 4\texttimes{} default model, and 4\texttimes{} large model
    \item BasicVSR++~\cite{chan2022basicvsrpp}: 4\texttimes{} model
    \item IART~\cite{xu2024enhancingvideosuperresolutionimplicit}: 4\texttimes{} model
    \item Topaz~\cite{topaz_video_ai}: 2\texttimes{} and 4\texttimes{} models; NYX preset
\end{itemize}
This procedure produced 1280 upscaled video clips in total.

Following this, we used \url{https://subjectify.us} to conduct a crowd-sourced pairwise comparison among each of the upsampled versions for every video clip.
In the comparison, participants were shown pairs of upsampling results, and asked to choose, which of the two they consider higher quality.
The comparison included control pairs to ensure accurate responses from participants.
In total, we collected 159,972 pairwise votes from 6153 participants.
We used the Bradley-Terry model~\cite{bradley1952rank} to aggregate the votes into a ground-truth rank for every upsampled video.

To improve the variety and coverage of video content, we extended our dataset with 2~videos from the MSU Video Super-Resolution Benchmark~\cite{kirillova2022erqa}, 4~videos from the MSU Super-Resolution for Video Compression Benchmark~\cite{srcodec}, and 2~videos from the MSU Video Upscalers Benchmark~\cite{videoupscalersbenchmark}, along with their compressed and upscaled variants (685 clips in total).
These benchmarks collected videos and conducted pairwise comparisons using a similar methodology to ours, with a different selection of codecs and SR models.

In order to fairly shuffle these distorted videos into train and test subsets, we further split them into 20 clusters, based on existing metrics' performance when applied to them.
This step used the following metrics:
\begin{itemize}
    \item Noise Estimation (from MSU VQMT~\cite{vqmt})
    \item Blurring (from MSU VQMT~\cite{vqmt})
    \item DBCNN~\cite{8576582}
    \item ClipIQA+~\cite{wang2022exploring}
    \item ERQA~\cite{kirillova2022erqa}
    \item LPIPS (VGG)~\cite{8578166}
    \item Q-Align (image quality assessment task)~\cite{wu2023qalign}
    \item HyperIQA~\cite{Su_2020_CVPR}
    \item TOPIQ (no-reference version)~\cite{chen2023topiq}
\end{itemize}
At this step, 841 videos were removed based on unusually good performance of existing metrics (too easy).
The remaining 1124 videos were split into train, public test and private test evenly across clusters.

\begin{figure}[tb]
    \centering
    \begin{subfigure}{0.31\textwidth}
        \includegraphics[width=\linewidth]{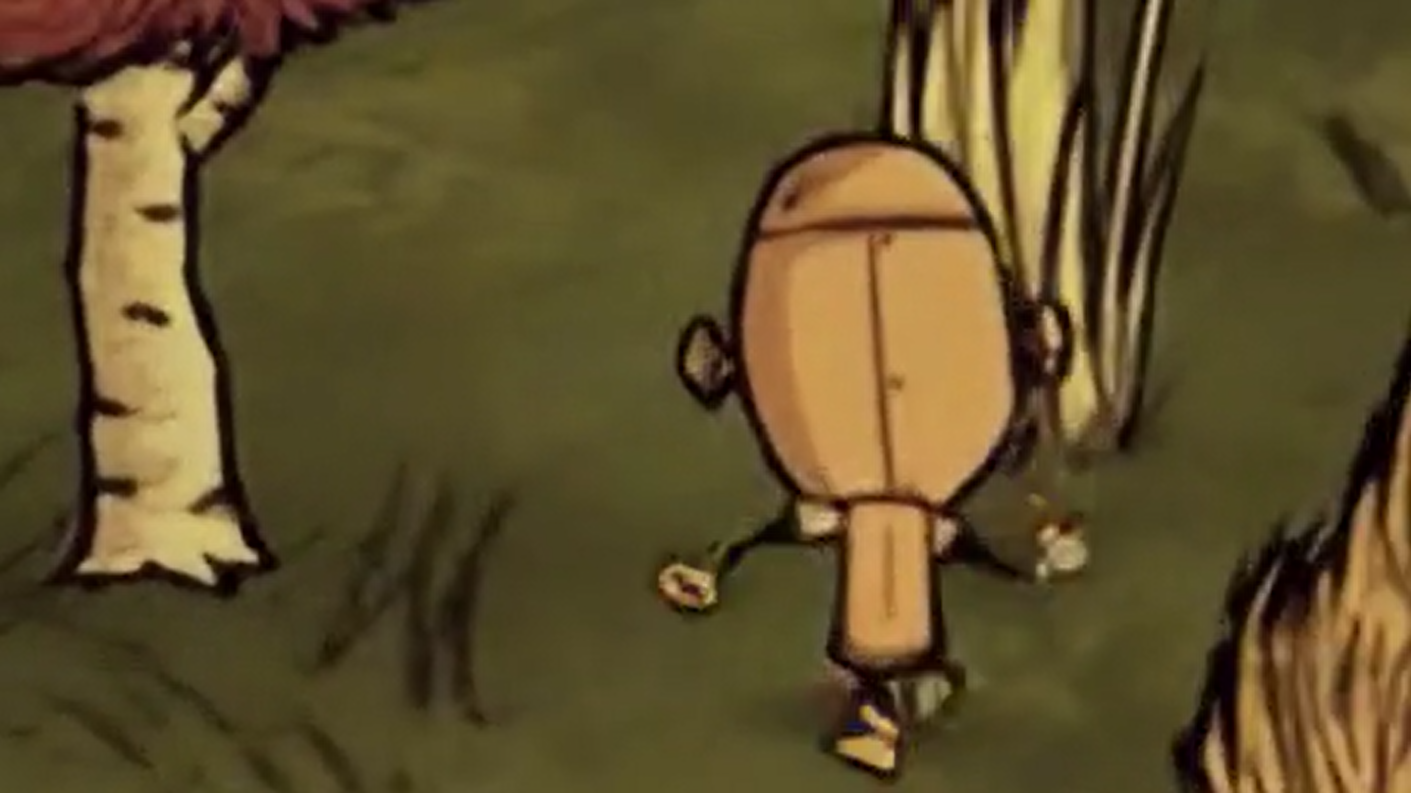}
        \caption{Easy Sample Clip}
    \end{subfigure}
    \begin{subfigure}{0.31\textwidth}
        \includegraphics[width=\linewidth]{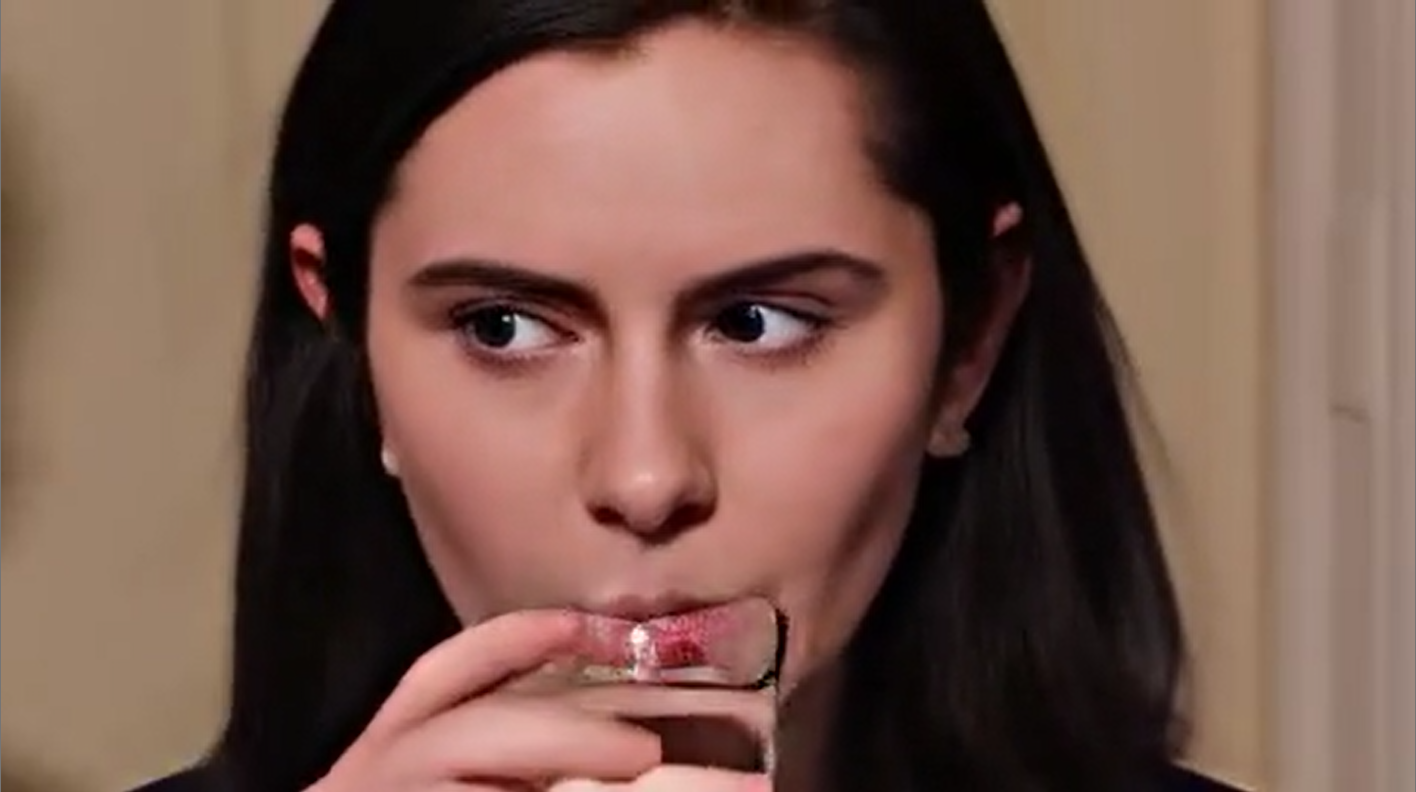}
        \caption{Medium Sample Clip}
    \end{subfigure}
    \begin{subfigure}{0.31\textwidth}
        \includegraphics[width=\linewidth]{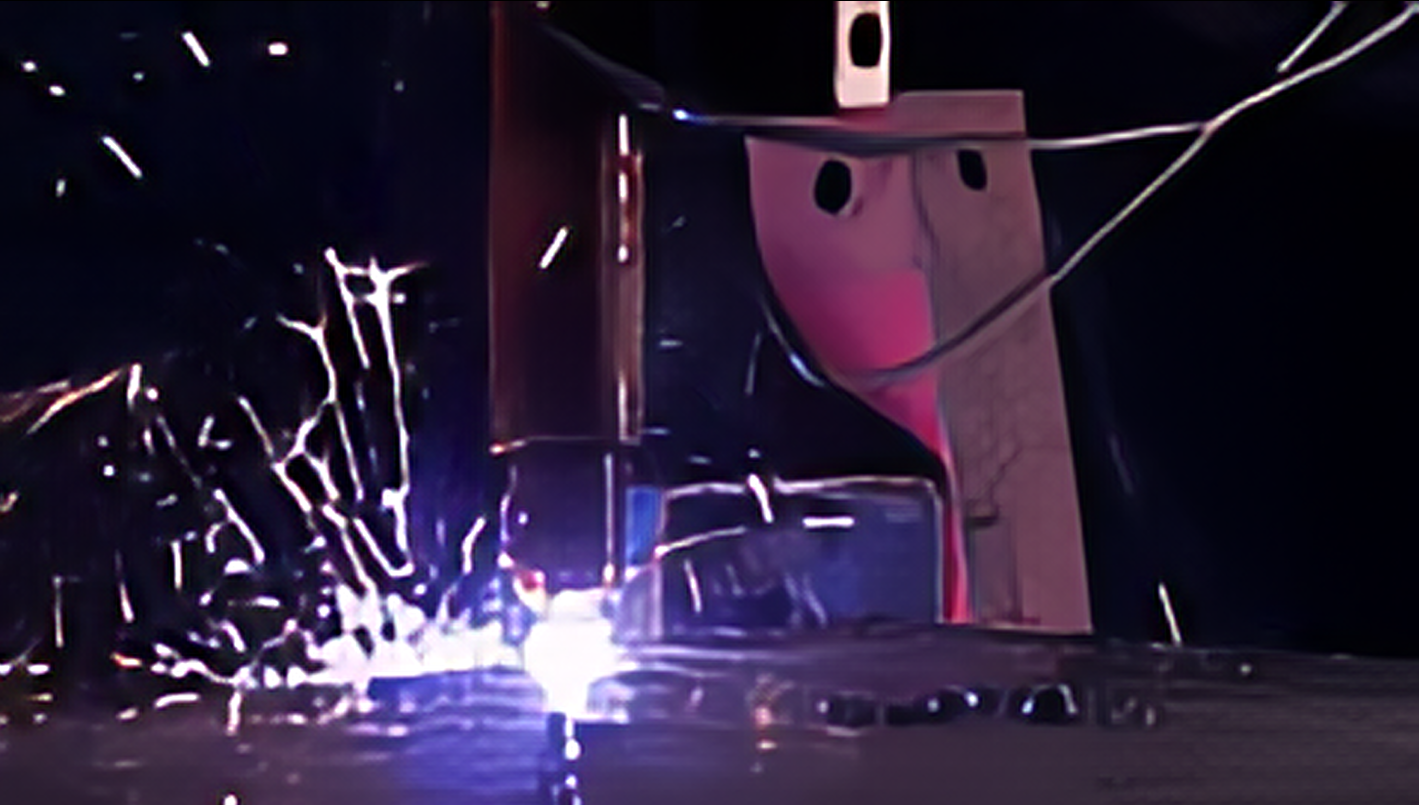}
        \caption{Hard Sample Clip}
    \end{subfigure}
    \caption{Sample clips from our public testing set for each difficulty level.}
    \label{fig:dataset}
\end{figure}

The final step was to split videos among three difficulty levels.
We categorized videos as Easy, Medium or Hard based on the performance of existing metrics according to our evaluation methodology described in~\cref{sec:evaluation}.
\Cref{fig:dataset} shows sample clips from the public testing set for each of the difficulty level.

Reviewing the final categorization, we make the following observations:
\begin{itemize}
    \item ``Easy'' level includes videos without special artifacts (or with very weak distortions): only blur, noise, etc.
    \item ``Medium'' level includes videos without special artifacts, videos with weak distortions, as well as videos with very strong SR distortions that are nevertheless caught by a significant part of metrics.
    \item ``Hard'' level includes videos with obvious distortions that are not handled by most metrics.
\end{itemize}

While the difficulty level was specified in our data, participants were not allowed to use it as input to their model, or to select models based on the difficulty level.
This is because it is specific to our dataset collection procedure, and cannot generally be computed independently for novel videos, while the goal of our challenge is to develop a generally applicable SR QA metric.

\subsection{Evaluation Protocol}
\label{sec:evaluation}

Our data collection process provides us with ground-truth subjective ranks among all upscaled versions of every individual video.
However, ranks are not directly comparable between clips of different source videos.
We therefore design our evaluation procedure to measure how well a given quality assessment method ranks upscaled clips from the same video.

\begin{table}[tb]
  \caption{Combined challenge results. Public Score: score on the public test set. Private Score: score on the private test set. Final Score: combined challenge score. Sorted by Final Score. The best result is \textbf{bold}, the second-best result is \underline{underlined}.}
  \label{tab:combined-results}
  \centering
\begin{tabular}{@{} l l c c c @{}}
  \toprule
  Team & Type & Public Score$^\uparrow$ & Private Score$^\uparrow$ & Final Score$^\uparrow$ \\
  \midrule
  QA-FTE                                       & NR Video &            0.8661  &    \textbf{0.8575} &    \textbf{0.8604} \\
  TVQA-SR                                      & NR Video &    \textbf{0.8907} & \underline{0.8448} & \underline{0.8601} \\
  SJTU MMLab                                   & NR Video & \underline{0.8906} &            0.8362  &            0.8543  \\
  Wink                                         & NR Video &            0.8864  &            0.8014  &            0.8297  \\
  sv-srcb-lab                                  & NR Video &            0.7926  &            0.8432  &            0.8263  \\
  PieAPP~\cite{Prashnani_2018_CVPR} (baseline) & FR Image &            0.6971  &            0.8025  &            0.7674  \\
  Q-Align~\cite{wu2023qalign} (baseline)       & NR Image &            0.7028  &            0.7855  &            0.7580  \\
  \bottomrule
\end{tabular}
\end{table}

For every source video, we compute Spearman's rank order correlation coefficient~\cite{srocc} between scores predicted by the participant's QA method for all upscaled versions of that video, and the ground-truth subjective ranks.
Then, we average these correlation coefficients among all source videos in each of the difficulty levels (Easy, Medium, Hard).
The final score for a test set is a weighted combination of these scores:
\begin{equation}
    \mathit{Score} = \frac{0.3 \cdot \mathit{Easy} + 0.4 \cdot \mathit{Medium} + 0.5 \cdot \mathit{Hard}}{0.3 + 0.4 + 0.5}.
\end{equation}

Over the course of the challenge, participants submitted their predicted labels for all video clips into our testing system.
The testing system computed scores for the public test set and uploaded them to the challenge's web page.
At the end of the challenge, we suspended the automatic testing system and asked participants to send in the final results and the model code and weights that reproduce these results.
We then ran each of the final models and verified that the results match those sent in by the teams.

We computed the scores for the private test set using the same procedure as for the public test set. The final score is a combination of the public and the private scores:
\begin{equation}
    \mathit{Final} = \frac13 \left(\mathit{Public} + 2 \cdot {Private}\right).
\end{equation}

\section{Challenge Results}
\label{sec:results}

\begin{table}[tb]
  \caption{Challenge results on the \textbf{private} test set. Easy, Medium, Hard: average of Spearman correlations across all videos from the respective subset. Score: described in \cref{sec:evaluation}. Sorted by the combined Final Score. The best result is \textbf{bold}, the second-best result is \underline{underlined}.}
  \label{tab:private-results}
  \centering
\begin{tabular}{@{} l l c c c c @{}}
  \toprule
  Team & Type & Easy$^\uparrow$ & Medium$^\uparrow$ & Hard$^\uparrow$ &  Private Score$^\uparrow$ \\
  \midrule
  QA-FTE                                       & NR Video &            0.8595  &    \textbf{0.9323} &    \textbf{0.7965} &    \textbf{0.8575} \\
  TVQA-SR                                      & NR Video &            0.8741  &            0.9115  &            0.7738  & \underline{0.8448} \\
  SJTU MMLab                                   & NR Video &    \textbf{0.9044} & \underline{0.9255} &            0.7239  &            0.8362  \\
  Wink                                         & NR Video &            0.8600  &            0.8986  &            0.6885  &            0.8014  \\
  sv-srcb-lab                                  & NR Video &            0.8758  &            0.9014  & \underline{0.7769} &            0.8432  \\
  PieAPP~\cite{Prashnani_2018_CVPR} (baseline) & FR Image &            0.8471  &            0.8820  &            0.7120  &            0.8025  \\
  Q-Align~\cite{wu2023qalign} (baseline)       & NR Image & \underline{0.8864} &            0.8456  &            0.6770  &            0.7855  \\
  \bottomrule
\end{tabular}
\end{table}

\begin{table}[tb]
  \caption{Challenge results on the \textbf{public} test set. Easy, Medium, Hard: average of Spearman correlations across all videos from the respective subset. Score: described in \cref{sec:evaluation}. Sorted by the combined Final Score. The best result is \textbf{bold}, the second-best result is \underline{underlined}.}
  \label{tab:public-results}
  \centering
\begin{tabular}{@{} l l c c c c @{}}
  \toprule
  Team & Type & Easy$^\uparrow$ & Medium$^\uparrow$ & Hard$^\uparrow$ &  Public Score$^\uparrow$ \\
  \midrule
  QA-FTE                                       & NR Video &            0.8899  &            0.8471  &            0.8669  &            0.8661  \\
  TVQA-SR                                      & NR Video &            0.9245  &            0.8763  &    \textbf{0.8819} &    \textbf{0.8907} \\
  SJTU MMLab                                   & NR Video &    \textbf{0.9383} &    \textbf{0.8832} & \underline{0.8679} & \underline{0.8906} \\
  Wink                                         & NR Video & \underline{0.9311} & \underline{0.8769} &            0.8672  &            0.8864  \\
  sv-srcb-lab                                  & NR Video &            0.8967  &            0.7607  &            0.7556  &            0.7926  \\
  PieAPP~\cite{Prashnani_2018_CVPR} (baseline) & FR Image &            0.8278  &            0.6877  &            0.6263  &            0.6971  \\
  Q-Align~\cite{wu2023qalign} (baseline)       & NR Image &            0.8908  &            0.7070  &            0.5867  &            0.7028  \\
  \bottomrule
\end{tabular}
\end{table}

\Cref{tab:combined-results} shows the final AIM 2024 VSR QA results.
\Cref{tab:private-results,tab:public-results} show score breakdown on the private and the public test sets, respectively.
Baseline image-QA metrics were computed frame-by-frame and averaged to give the combined score for the video.

The QA-FTE team has won the first place on both our private test set and the challenge as a whole, while the TVQA-SR team took the second place overall while showing the best result on our public test set.

\subsection{Result Analysis}
\label{sec:analysis}

Every team has confident wins over the baselines on the public test set with TVQA-SR and SJTU MMLab taking the lead, and Wink showing good scores on the Easy and Medium levels.
The private set has the results mixed up a bit with QA-FTE having an advantage on both Medium and Hard levels, sv-srcb-lab showing second-best Hard performance, while the Q-Align baseline takes second place on the Easy level.
SJTU MMLab has very good Easy and Medium scores on both test sets, suggesting that their architecture should be a great fit for a wide range of typical video content.

All participants opt for a no-reference video metric approach, combining various per-frame and inter-frame features.
Q-Align~\cite{wu2023qalign} in particular makes a recurring appearance in 4 out of 5 submissions indicating its good image quality assessment performance.
Among inter-frame features, SlowFast~\cite{feichtenhofer2019slowfast} seems to be a strong contender, also used in 4 out of 5 submissions.
Fast-VQA~\cite{wu2022fast}, Mamba~\cite{vim}, Swin-B~\cite{liu2021swin}, and ConvNeXt-V2~\cite{Woo_2023_CVPR} were each used in 2 submissions.

\section{Challenge Methods and Teams}
\label{sec:methods}

In this section, each team briefly describes their solution.
Teams appear in order of the final ranking.

\subsection{QA-FTE}

\begin{figure}[tb]
    \centering
    \includegraphics[width=\textwidth]{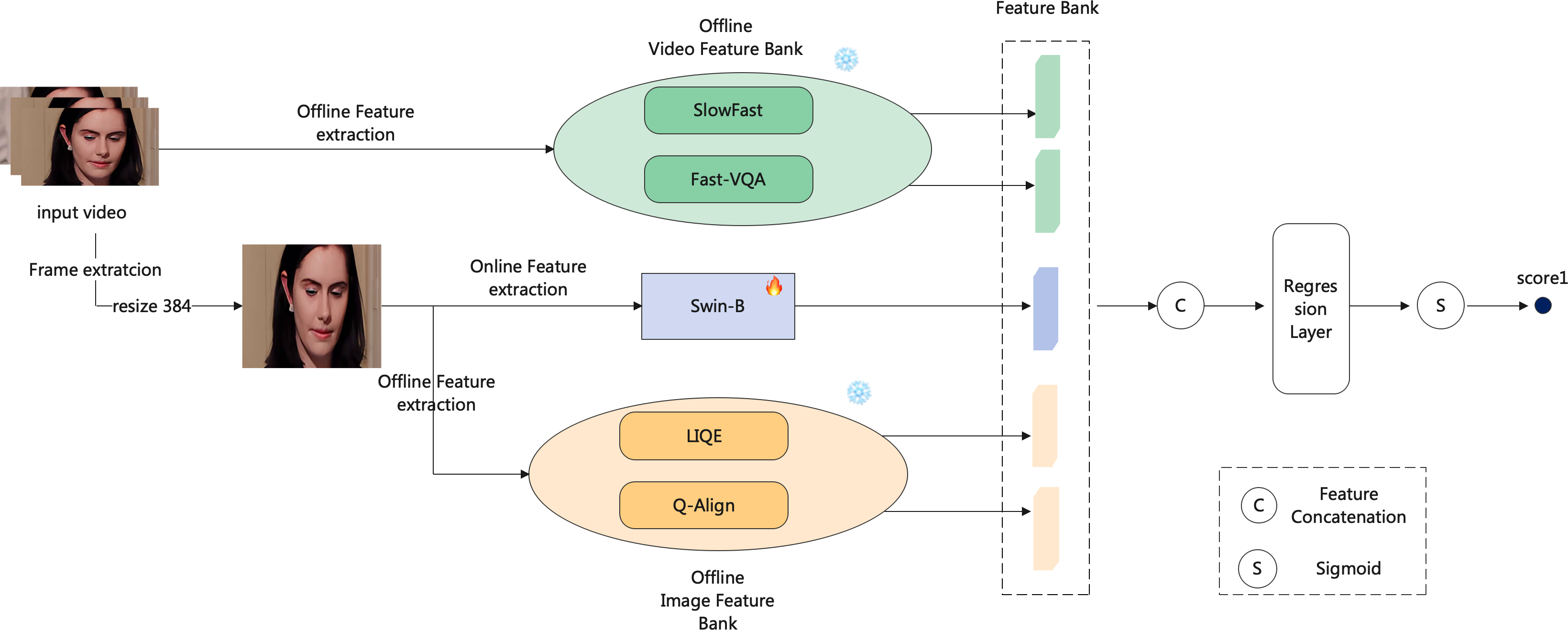}
    \caption{QA-FTE architecture pipeline.}
    \label{fig:QA-FTE}
\end{figure}

The overall architecture of our method is shown in~\cref{fig:QA-FTE}.
Given the diverse visual content and the complex distortion types, we exploit rich features to equip the model with stronger generalization ability, following the idea of previous works~\cite{sun2024enhancing,wang2021rich}.
Swin Transformer-B~\cite{liu2021swin,liu2022swiniqa} which is pretrained on LSVQ dataset~\cite{ying2021patch} is adopted as the backbone for learning spatial quality feature representation. The offline video feature bank provides temporal and spatial-temporal feature representations, coming from SlowFast~\cite{feichtenhofer2019slowfast} and Fast-VQA~\cite{wu2022fast} respectively.  The  offline image feature bank provides comprehensive frame-level  
feature representaions, where LIQE~\cite{zhang2023blind} contains quality-aware, distortion-specific as well as scene-specific information, and Q-Align\cite{sammeth2003qalign} contains strong quality-aware
features benefiting from large multi-modality models. The learnable and non-learnable features are concatenated together to predict the final score, which is finally converted to the range of [0-1] by the Sigmoid function.

\begin{figure}[tb]
    \centering
    \includegraphics[width=0.6\textwidth]{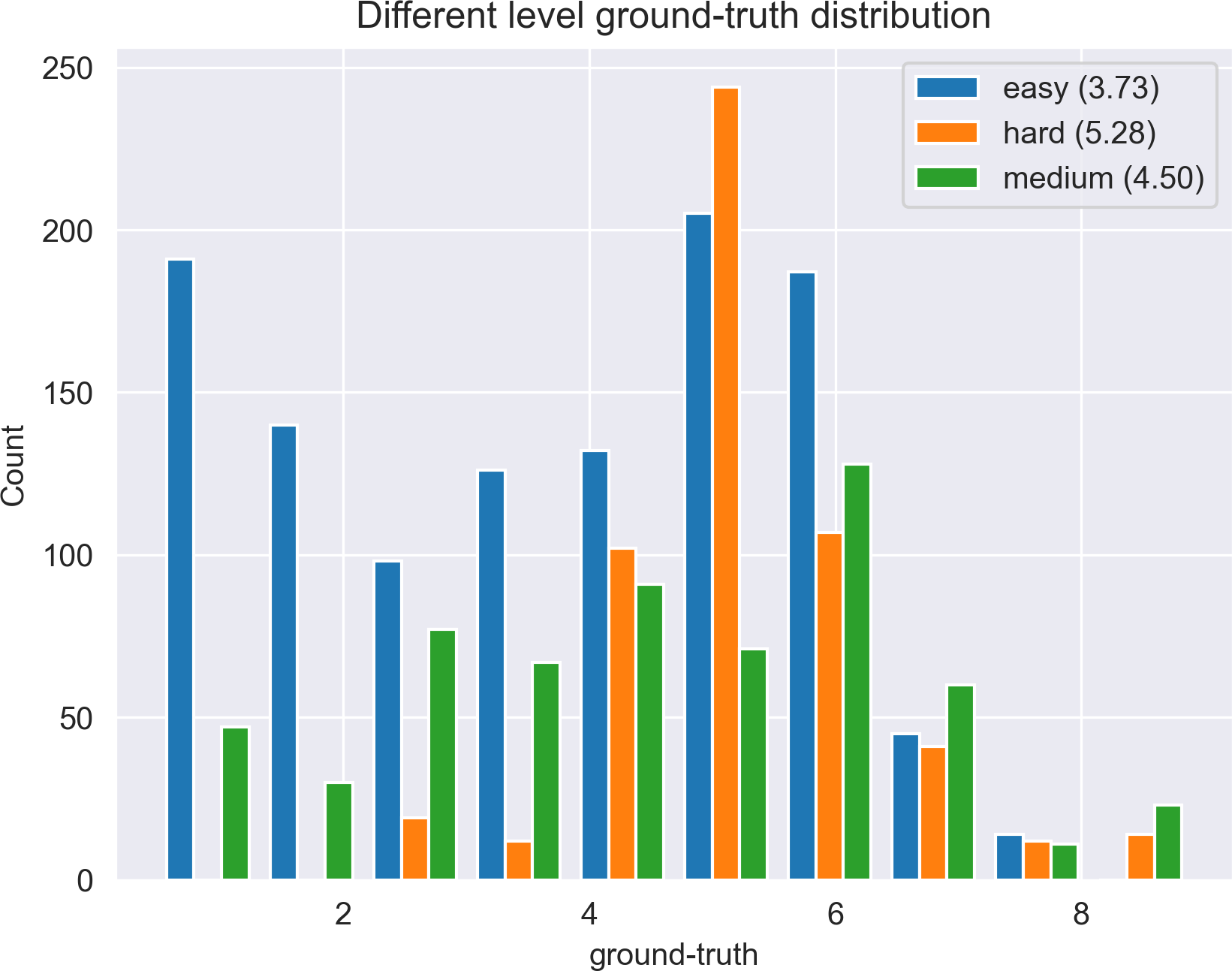}
    \caption{Distributions of subjective scores.}
    \label{fig:QA-FTE_lvl_dif2}
\end{figure}

\begin{figure}[tb]
    \centering
    \includegraphics[width=\textwidth]{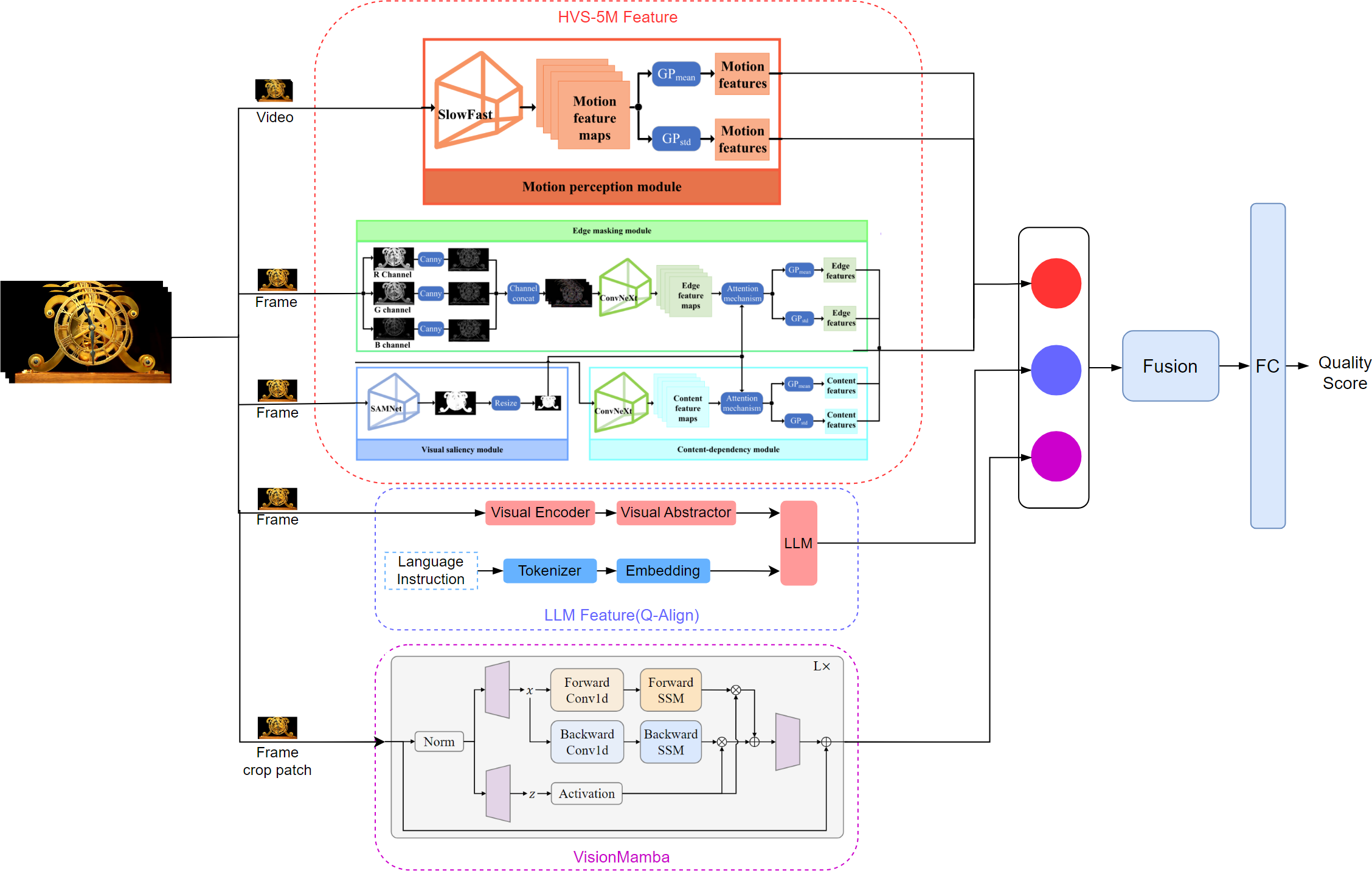}
    \caption{Architecture of TVQA-SR.}
    \label{fig:tvqa_sr}
\end{figure}

\begin{figure}[tb]
    \centering
    \includegraphics[width=\textwidth]{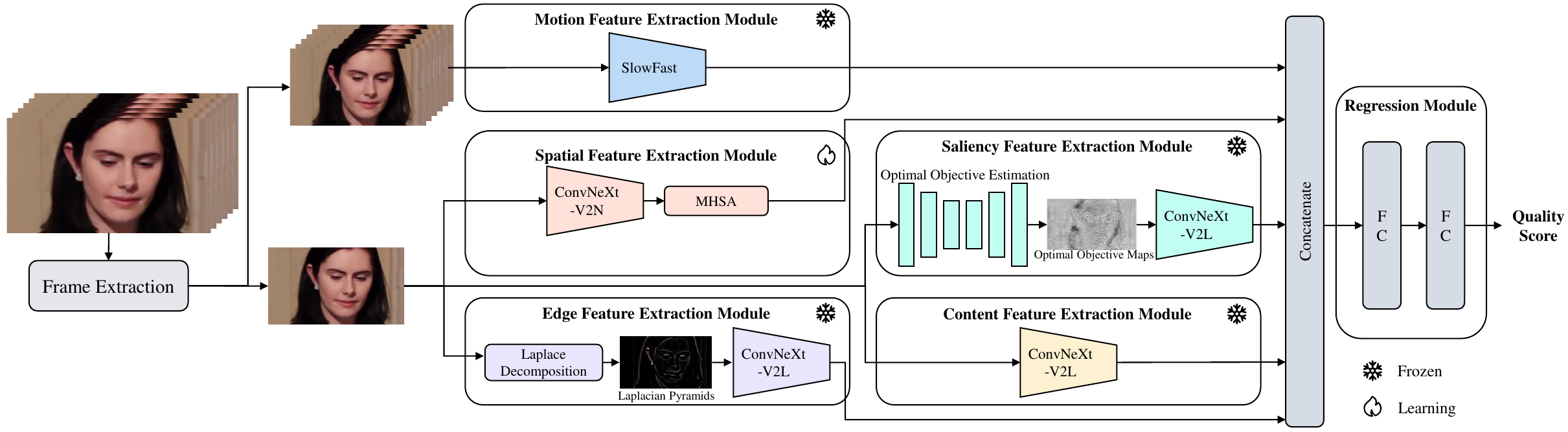}
    \caption{The framework of our proposed SR-VQA model.}
    \label{fig:SJTUMMLab}
\end{figure}

\begin{figure}[tb]
    \centering
    \includegraphics[width=.8\textwidth]{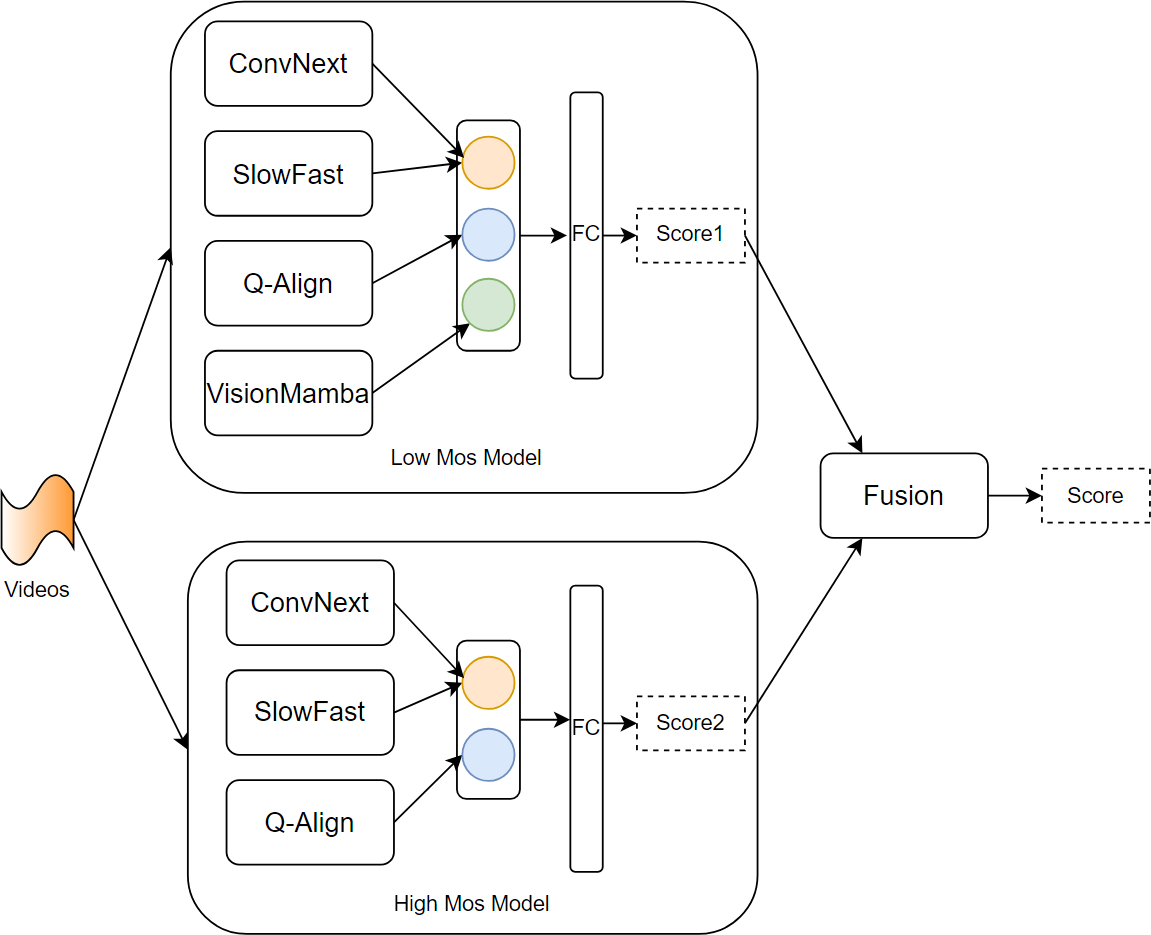}
    \caption{FusionVQA architecture.}
    \label{fig:wink}
\end{figure}

\subsubsection{Training Details} We analyze the distribution of subjective scores for each group. From~\cref{fig:QA-FTE_lvl_dif2} we can see that the subjective scores of hard group videos are harder to distinguish compared with easy and medium groups. Therefore,  apart from PLCC loss~\cite{sun2024enhancing}, we also apply the pairwise ranking hinge loss~\cite{liu2017rankiqa} to  guide the model to distinguish the hard samples while quickly learning the easy samples. The training loss is:
\begin{equation}
    {L}={{L}_{Rank}} + 2.0 * {{L}_{PLCC}} ,
\end{equation}
where the rank margin is set to 0.05. We train the model with a learning rate of $1e^{-5}$ for 100 epochs on 8 A100-SXM-80GB GPUs with a batch size of 16. We randomly sampled 80\% videos from the training data for training, 20\% for validation, and choose the model with the best validation performance on hard group as the final model.

\subsection{TVQA-SR}

We use HVS-5M~\cite{HVS_5M} to extract CNN-based video spatial features and motion features. Subsequently, Q-Align~\cite{wu2023qalign} is utilized to extract features from the video frame to enhance the semantic expression ability of the features, and VisionMamba~\cite{vim} is used to extract quality features from the crop patch of each frame. Then a feature fusion module is adopted to fuse the extracted features mentioned above. And finally they are passed through a FC layer to obtain the quality score. The model architecture is shown in~\cref{fig:tvqa_sr}.

During the training phase, PLCC loss and SROCC loss are used, and we train the model on Nvidia V100 GPU with a batch size of 32 for 100 epochs with a learning rate of 0.00005.

\subsection{SJTU MMLab}

We propose the Super-Resolution Video Quality Assessment (SR-VQA) method, based on UNQA~\cite{cao2024unqa}, which comprises a SlowFast for motion feature extraction, a ConvNeXt-V2N for spatial feature extraction, and a ConvNext-V2L for edge, saliency, and content feature extraction. The whole framework is shown in \cref{fig:SJTUMMLab}. Edged features are extracted from Laplacian pyramids derived from the key frames. We utilize the predictive model of SISR networks~\cite{park2023perception} to generate the optimal objective maps and extract saliency features. Finally, we concatenate these features to predict the video quality scores via two-layer MLP.

\subsubsection{Training Details}

The weights of the spatial feature extraction module are pretrained on four IQA databases (BID~\cite{ciancio2010no}, CLIVE~\cite{ghadiyaram2015massive}, KonIQ10K~\cite{hosu2020koniq}, and SPAQ~\cite{fang2020perceptual}) and four VQA databases (LSVQ~\cite{ying2021patch}, YouTube-UGC~\cite{wang2019youtube}, KoNViD-1k~\cite{hosu2017konstanz}, and LIVE-VQC~\cite{sinno2018large}). For the motion feature extraction module, the resolution of the videos is resized to $224\times 224$ and videos are split into several $1$~s video chunks as inputs. The other modules sample one key frame from $1$~s video chunks. For the spatial feature extraction module, the resolution of the videos is resized to $384\times 384$. For the saliency, edge, and content feature extraction module, the original resolution videos are used as inputs. We train the proposed model on Nvidia RTX 4090 GPU with a batch size of $6$ for $50$ epochs. The learning rate is set as $0.00001$.

\subsection{Wink}

We proposed FusionVQA, which train High Mos Model and Low Mos Model for high-quality data and low-quality data respectively. When predicting, the scores of the two models are fused according to the MOS score predicted by the High Mos Model. Score=Score2 if Score2>2, otherwise Score=Score1. \Cref{fig:wink} shows an overview of our method's architecture.

\subsection{sv-srcb-lab}

\begin{figure}[tb]
    \centering
    \includegraphics[width=.8\textwidth]{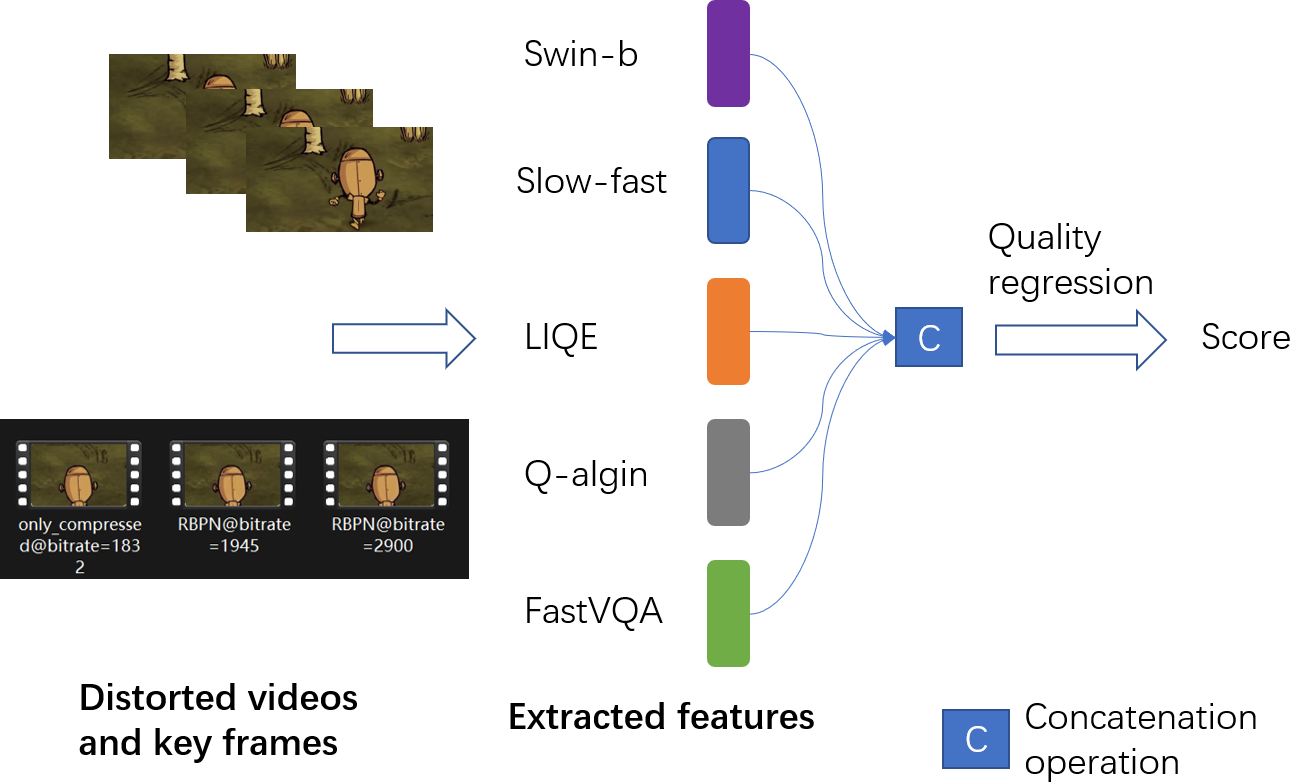}
    \caption{Architecture pipeline of sv-srcb-lab.}
    \label{fig:sv-srcb-lab}
\end{figure}

We basically follow the pipeline of the paper~\cite{sun2024enhancing} with little modification. The architecture pipeline is shown in \cref{fig:sv-srcb-lab}. We first extract 5 different features from the input key frames and videos, then fuse and feed them to the quality regression module, which outputs a quality score. During the training, the 568 train videos were randomly split into 454 and 114 clips for train and validation, respectively, without dataset enlarge applied.

\section{Conclusion}

This paper presented our AIM 2024 challenge on video super-resolution quality assessment.
To ensure a fair evaluation, we have collected a diverse dataset of 1124 videos upscaled using 52 modern SR methods.
The ground-truth ranks were obtained by a crowd-source pairwise comparison with >150,000 votes.

The challenge had 5 teams make the final submission, and each of the submitted methods surpassed the current state-of-the-art in image- and video-QA, with the QA-FTE team taking the first place.
All metrics are no-reference, and are based on combinations of deep-learning-based per-frame and inter-frame features.

We look forward to the future development of the field of SR QA.
One interesting next step would be to evaluate metrics on challenging cases with artifacts produced by SR methods.

\section*{Acknowledgements}
This work was partially supported by the Humboldt Foundation. We thank the AIM 2024 sponsors: Meta Reality Labs, KuaiShou, Huawei, Sony Interactive Entertainment and University of W\"urzburg (Computer Vision Lab).

\appendix
\section{Teams and Affiliations}
\label{affiliations}

\subsection{QA-FTE}

\textit{\textbf{Members:}}\\
Jianzhao Liu$^1$ (\nolinkurl{liujianzhao.0622@bytedance.com}), Tianwu Zhi$^1$ (\nolinkurl{zhitianwu@bytedance.com}), Yabin Zhang$^1$, Yang Li$^1$, Jingwen Xu$^1$, Yiting Liao$^1$\\
\textit{\textbf{Affiliations:}}\\
$^1$: ByteDance, China

\subsection{TVQA-SR}

\textit{\textbf{Members:}}\\
Qing Luo$^{*,1}$ (\nolinkurl{luoqing.94@qq.com}), Ao-Xiang Zhang$^{*,1,2}$ (\nolinkurl{zax@e.gzhu.edu.cn}), Peng Zhang$^1$, Haibo Lei$^1$, Linyan Jiang$^1$, Yaqing Li$^1$\\
$^*$: Equal contribution.\\
\textit{\textbf{Affiliations:}}\\
$^1$: Tencent, China\\ 
$^2$: School of Computer Science and Cyber Engineering, Guangzhou University, China

\subsection{SJTU MMLab}

\textit{\textbf{Members:}}\\
Yuqin Cao$^1$ (\nolinkurl{caoyuqin@sjtu.edu.cn}), Wei Sun$^1$ (\nolinkurl{sunguwei@sjtu.edu.cn}), Weixia Zhang$^{1}$ (\nolinkurl{zwx8981@sjtu.edu.cn}), Yinan~Sun$^{1}$ (\nolinkurl{yinansun@sjtu.edu.cn}), Ziheng Jia$^{1}$ (\nolinkurl{jzhws1@sjtu.edu.cn}), Yuxin Zhu$^1$(\nolinkurl{rye2000@sjtu.edu.cn}), Xiongkuo Min$^{1}$ (\nolinkurl{minxiongkuo@sjtu.edu.cn}), Guangtao Zhai$^{1}$ (\nolinkurl{zhaiguangtao@sjtu.edu.cn})\\
\textit{\textbf{Affiliations:}}\\
$^1$: Shanghai Jiao Tong University, Shanghai, China

\subsection{Wink}

\textit{\textbf{Members:}}\\
Weihua Luo$^1$ (\nolinkurl{185471613@qq.com})\\
\textit{\textbf{Affiliations:}}\\
$^1$: None, China

\subsection{sv-srcb-lab}

\textit{\textbf{Members:}}\\
Yupeng Z.$^1$, Hong Y.$^1$\\
\textit{\textbf{Affiliations:}}\\
$^1$: Ricoh Software Research Center Beijing, China

\subsection{Challenge Organizers}

\textit{\textbf{Members:}}\\
Ivan Molodetskikh$^{*,1}$ (\email{ivan.molodetskikh@graphics.cs.msu.ru}),\\
Artem Borisov$^{*,1}$ (\email{artem.borisov@graphics.cs.msu.ru}),\\
Dmitriy Vatolin$^{1,2}$ (\email{dmitriy@graphics.cs.msu.ru}),\\
Radu Timofte$^3$ (\email{radu.timofte@uni-wuerzburg.de})\\
$^*$: Equal contribution.\\
\textit{\textbf{Affiliations:}}\\
$^1$: Lomonosov Moscow State University, Russia\\
$^2$: MSU Institute for Artificial Intelligence, Russia\\
$^3$: University of Würzburg, Germany

%
%
\bibliographystyle{splncs04}
\bibliography{main}
\end{document}